\documentclass[runningheads]{llncs}
\usepackage{graphicx}
\begin{document}
\title{Recognition and Localisation of Pointing Gestures using a RGB-D Camera}
%
%
\author{Naina Dhingra\orcidID{0000-0001-7546-1213}
\and
Eugenio Valli
\and
\\Andreas Kunz\orcidID{0000-0002-6495-4327}}
\authorrunning{N. Dhingra et al.}
%
\institute{
Innovation Center Virtual Reality, ETH Zurich, Zurich, Switzerland\\
\email{\{ndhingra,kunz\}@iwf.mavt.ethz.ch}\\
\url{https://www.icvr.ethz.ch} }
\maketitle              
\begin{abstract}
Non-verbal communication is part of our regular conversation, and multiple gestures are used to exchange information. Among those gestures, pointing is the most important one. If such gestures cannot be perceived by other team members, e.g. by blind and visually impaired people (BVIP), they lack important information and can hardly participate in a lively workflow. Thus, this paper describes a system for detecting such pointing gestures to provide input for suitable output modalities to BVIP. Our system employs an RGB-D camera to recognize the pointing gestures performed by the users. The system also locates the target of pointing e.g. on a common workspace. We evaluated the system by conducting a user study with 26 users. The results show that the system has a success rate of 89.59 and 79.92 \% for a $2 \times 3$ matrix using the left and right arm respectively, and  73.57 and 68.99  \% for $3 \times 4$ matrix using the left and right arm respectively.

\end{abstract}

\section{\uppercase{Introduction}}
\label{sec:introduction}
In a meeting environments, when sighted people and BVIP are working together, sighted people tend to do some habitual gestures, from which the most common ones are: facial expressions, hand gestures, pointing gestures, eye gaze, etc. There are in total 136 gestures \cite{brannigan1972human} which are termed as part of non-verbal communication (NVC). They need to be understood together with the verbal communication to understand the complete meaning of the conversation. However, for BVIP, the information from visual gestures are missing  \cite{gunther2019mapvi}. To understand the meaning of pointing gestures, it is crucial to know where a person is pointing at. Pointing gestures are the most common ones in nonverbal communication, and they become important in meetings where the speakers point towards objects in the room, or at artefacts on a whiteboard, as a reference to their speech. However, these pointing gestures are not accessible for BVIP and thus they lack important information during a conversation within a team meeting. To address this issue, we developed a system that automatically detects pointing gestures and determines the position where a person is pointing at. However, although NVC is easily understood by the humans, it is difficult for machines to recognize and interpret it reliably \cite{dhingra2019res3atn} and to avoid false alerts to the BVIP.

The main contributions of this paper are as follows: (1) We developed an autonomous system using OpenPTrack and ROS (Robot Operating System) to detect and localise the position of a pointing gesture. (2) We designed our system to work in real time and performed experiments using $2 \times 3$ and $3 \times 4$ grids. (3) We conducted a user study with 26 users to evaluate our system. We expect that our work will help the researchers to integrate BVIP in team meetings.

This paper is organized as follows: Section 2 describes the state of the art in OpenPtrack software, pointing and related gestures. Section 3 describes the methods and techniques used in our system, while section 4 gives an overview of the user study conducted and the setup of the built system. Finally, section 5 discusses the results obtained from the user study as well as the accuracies obtained by our system in detecting and localizing pointing gestures.

\section{\uppercase{State of the art}}
\subsection{OpenPtrack}
OpenPTrack is an open source software for tracking people and calibrating a multi- RGB-D camera setup \cite{munaro2016openptrack}. It can track multiple people at a frame rate of the sensor. It can also employ heterogeneous 3D sensors group. OpenPTrack uses a calibration procedure which depends on ROS communication networking capabilities and communication. In the past, detection and tracking systems exploited the color and depth information of a user, since cheap RGB-D sensors are available. Further, previous software were limited to single camera usage tracking system. These systems did not use multiple cameras and could not be implemented in distributed settings. Four our system, we use OpenPtrack since it allows expanding our pointing gesture system to multiple camera setup.

\subsection{Pointing Gesture Recognition}
Pointing gestures can be measured in different ways. Glove based techniques were used initially to sense the gesture being performed by the hand \cite{quam1990gesture}. Nowadays, computer vision based techniques  \cite{dhingra2019res3atn,rautaray2015vision} or Hidden Markov Models (HMMs)  \cite{wilson1999parametric} are used for detection. In particular, for pointing gesture detection, cascaded HMMs along with a particle filter was used for pointing gesture detection in \cite{park2011real}. The HMM in their first stage takes estimation of the hand position. It maps the estimated position to a precise position by modeling the kinematic features of the pointing finger. The output 3D coordinates are fed into their HMM in a second stage that differentiates the pointing gestures from other types of gesture. This technique requires a long processing time and a large training dataset.

Deep learning \cite{lecun2015deep}  has been successfully used in various applications of computer vision, which has inspired its use for gesture and body pose estimations as well \cite{neverova2014multi}. Deep learning approaches also solved pointing gesture recognition in \cite{huang2016pointing}. However, it requires large training dataset and only works with the specific data type on which it was trained on.

Our problem statement is to solve the pointing gesture recognition for BVIP more robustly. Using deep learning approaches would have required a large training dataset to make them applicable on different setups, such as a variety of meeting room layouts with a different number of people interacting at the same time. Thus, we chose a traditional way by using mathematical geometry and feature localisation. At first, a Kinect sensor along with OpenPtrack is used to locate the body joints. Next a mathematical geometry transformation is applied to achieve the spatial position of the pointing gesture's target. This position is classified into 6 fields (for $2 \times 3$ matrix), and into 12 fields (for $3 \times 4$ matrix).

\section{\uppercase{Methodology}}

The implementation of our pointing gesture recognizer and localizing system is based on OpenPTrack \cite{carraro2018real}. Using Kinect v2 as sensor, this software allows person tracking in real time over large areas. However, since this framework is not capable of directly detecting pointing gestures or other behavioral features of a user, we also forward the data to ROS \footnote{https://www.ros.org/}. By doing so, we can obtain the joints' coordinates in space for human gestures such as pointing. The main idea is that different packages of ROS could be implemented that contain so-called nodes, which are units that perform logic and computation for different parts of a robot, i.e. control of actuators, transform or change resolution of images provided by a sensor, etc. The different nodes of ROS can communicate with each other in order to share useful information or the functioning of the whole system, which is done by the \textit{topics}. Every node implementation of a pointing gesture recognizer for blind users can subscribe to such a topic to receive information or publish on a topic to share its content. OpenPtrack thus uses ROS to allow the information provided by Kinect to be further processed. The joints x, y and z coordinates are published under a repository 
which also contains different IDs for the different joints. The coordinate transformation from the sensor's reference frames to the world reference frame are performed using a ROS package called /TF, which rotates and translates the reference frames to the desired positions.

\begin{figure}[!h]
  \centering
  \begin{tabular}{cc}
  \includegraphics[width=6cm,height=4.0cm]{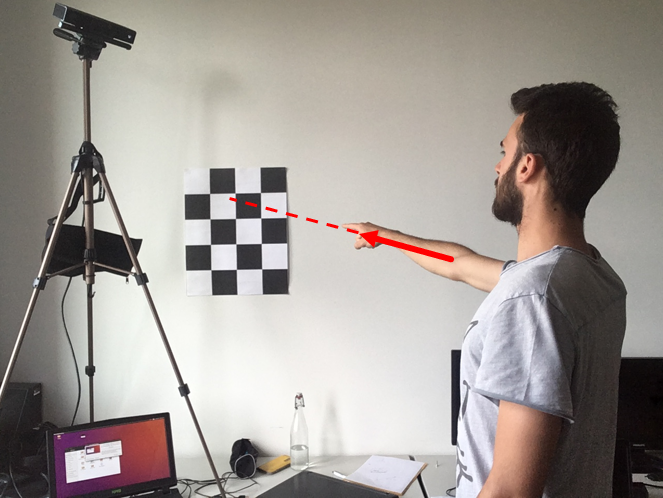}&
    \includegraphics[width=7cm,height=4.0cm]{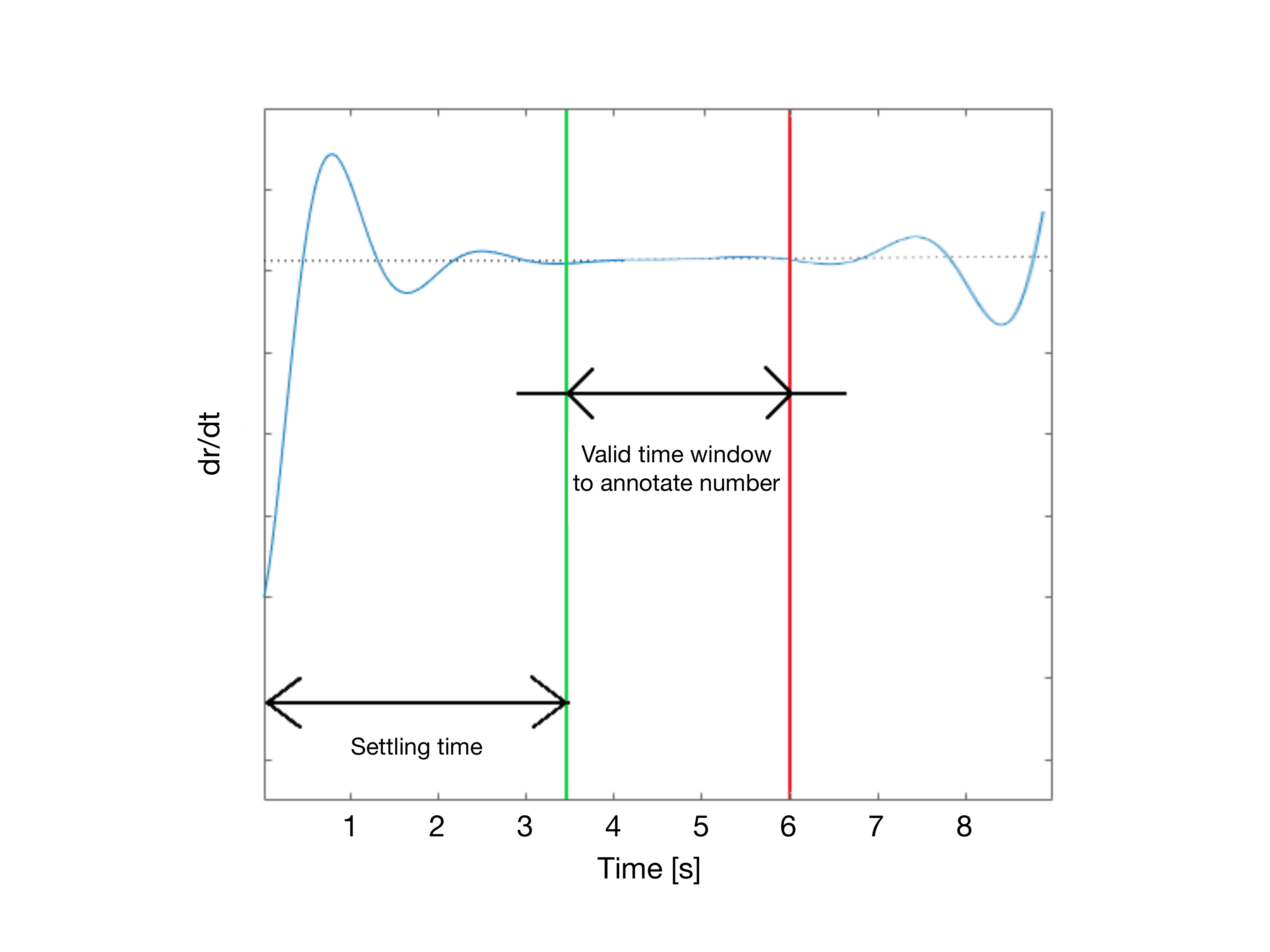}
     \end{tabular}
  \caption{Left:Pointing gesture with pointing vector. Right: Stabilization time of a pointing gesture where dr/dt is the change in the circle's diameter.}
     \label{fig:pointingvector}
 \end{figure}

A deictic (or pointing) gesture consists of the movement of an arm to point at a target in space and to highlight it by this gestures for other people  without necessarily having to verbally describe its position exactly. The joint's coordinates that have to be obtained are thus from elbow and hand, since these represent the human forearm and hence the major components for pointing. In order to define a pointing action, the link connecting the two aforementioned joints was measured, and named pointing vector as shown in the Figure \ref{fig:pointingvector}.

  Equation \ref{eq:pointing} is used to locate the position of the target on a vertical plane (e.g. a whiteboard) the user is pointing at.
\begin{equation}\label{eq:pointing}
    P_p=H+\frac{(H-P_f)\cdot\vec{N}_f}{\vec{EH}\cdot\vec{N}_f}\cdot\vec{EH},
\end{equation}

where $P_f$ is a point predefined on the ground plane, $\vec{N}_f$ is the normal vector to the plane and $H,\ E$ are the positions of hand and elbow joints, respectively.

OpenPtrack defines all measurements in a world reference frame. To understand the definition of the world reference frame in OpenPtrack, the TF package of ROS is used, which is a predefined package for coordinate transformation using rotation matrices and quaternions. The next step is to define a whiteboard/matrix plane coordinate frame in order to obtain the measured target point on it. This plane coordinate frame is achieved by applying the rotation matrix in relation world coordinate frame. The output values from OpenPtrack are converted in whiteboard/matrix plane coordinate frames. These converted values analysed by putting hard limits for each box in the matrix along both $x$ and $y$ direction. All of these values are evaluated on run time.

Before getting the information from the system, on which target a user is pointing at, it is required to wait about 3-3.5 sec for the pointing gestures to get stabilized. This waiting time is for a user to reach the stable pointing gesture without moving or vibrating his/her arm. The stability output from the system is achieved after the setting time as shown in Figure \ref{fig:pointingvector}. It also has to be noted that the pointing gesture will become unstable again after certain time period.


\section{\uppercase{Experimental Setup}}
The setup resembles an environment in which sighted users have to perform pointing gestures, which are automatically recognized by our system. The pointing gesture's target will then be determined to be provided to a suitable output device for BVIP. The experiments consist of four parts: two studies using the left arm and two using the right arm for pointing. The pointing gestures have to be performed on two different grid sizes in order to evaluate the accuracy of our system, i.e., at each $2 \times 3$ and $3 \times 4$ grid printed on the board.
 
\begin{figure}[!h]
  \centering
  \begin{tabular}{cc}
   \includegraphics[width=4.0cm,height=5.5cm]{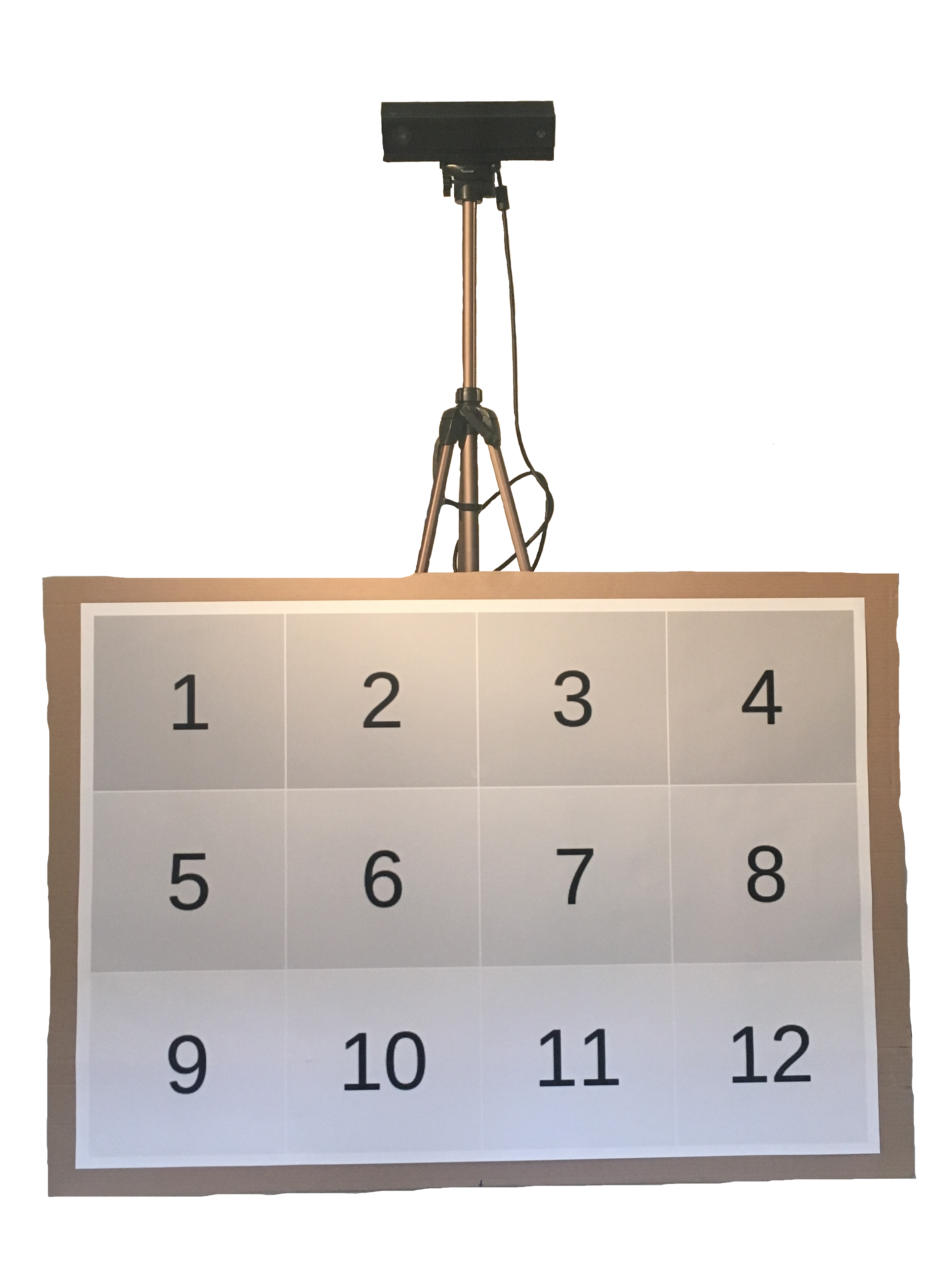}&
      \includegraphics[width=4.0cm,height=5.5cm]{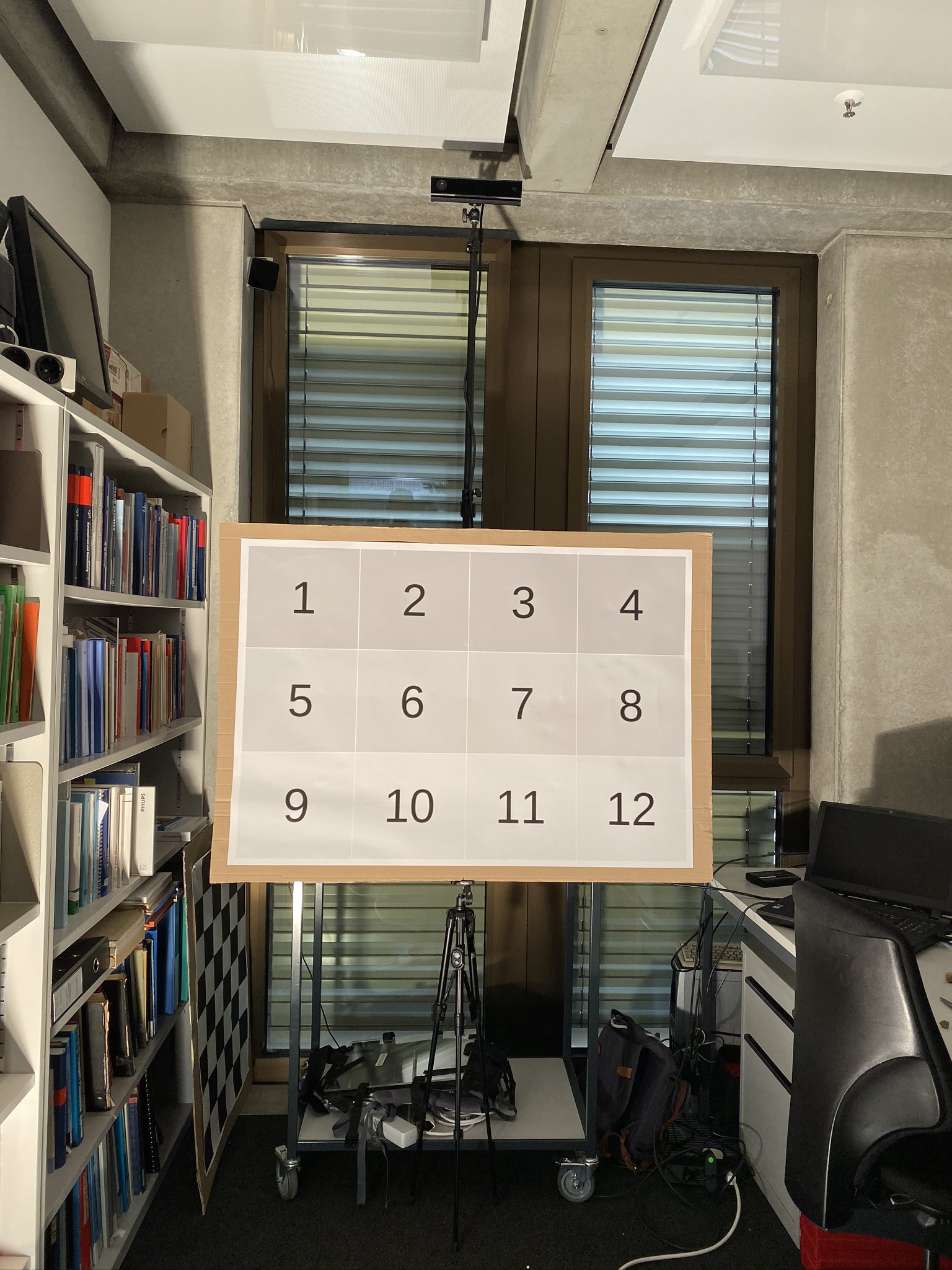}
   \end{tabular}
  \caption{Left: Measurement setup; Right: Experimental setup of the system. The Kinect is placed above the board having the matrix of numbers for the user to point at.}
  \label{fig:setup}
 \end{figure}

 The setup is shown in Figure \ref{fig:setup}. The board has the dimensions 1290mm x 1910mm and was 1000mm above the ground. The Kinect sensor was placed at a height of 1300mm above the top edge of the board and centered. Each box in the grid was numbered. The setup is shown in \ref{fig:setup}. The user had to stand at a constant distance of 1.5 m and centered in front of the board. Then he was asked to point towards the numbers following a given sequence told by the experimenter, and point for a few seconds to achieve a stable gesture before moving to the next number in the sequence. The stability time procedure is illustrated in Figure \ref{fig:pointingvector}. 
After the user was prompted to point at a certain box and the wait-time from Figure \ref{fig:pointingvector} was exceeded, the measured target number was recorded.

\section{\uppercase{User Study and Results}}

\begin{figure}[!htp]
  \centering
    \begin{tabular}{cc}
   \includegraphics[width=5.5cm,height=4.0cm]{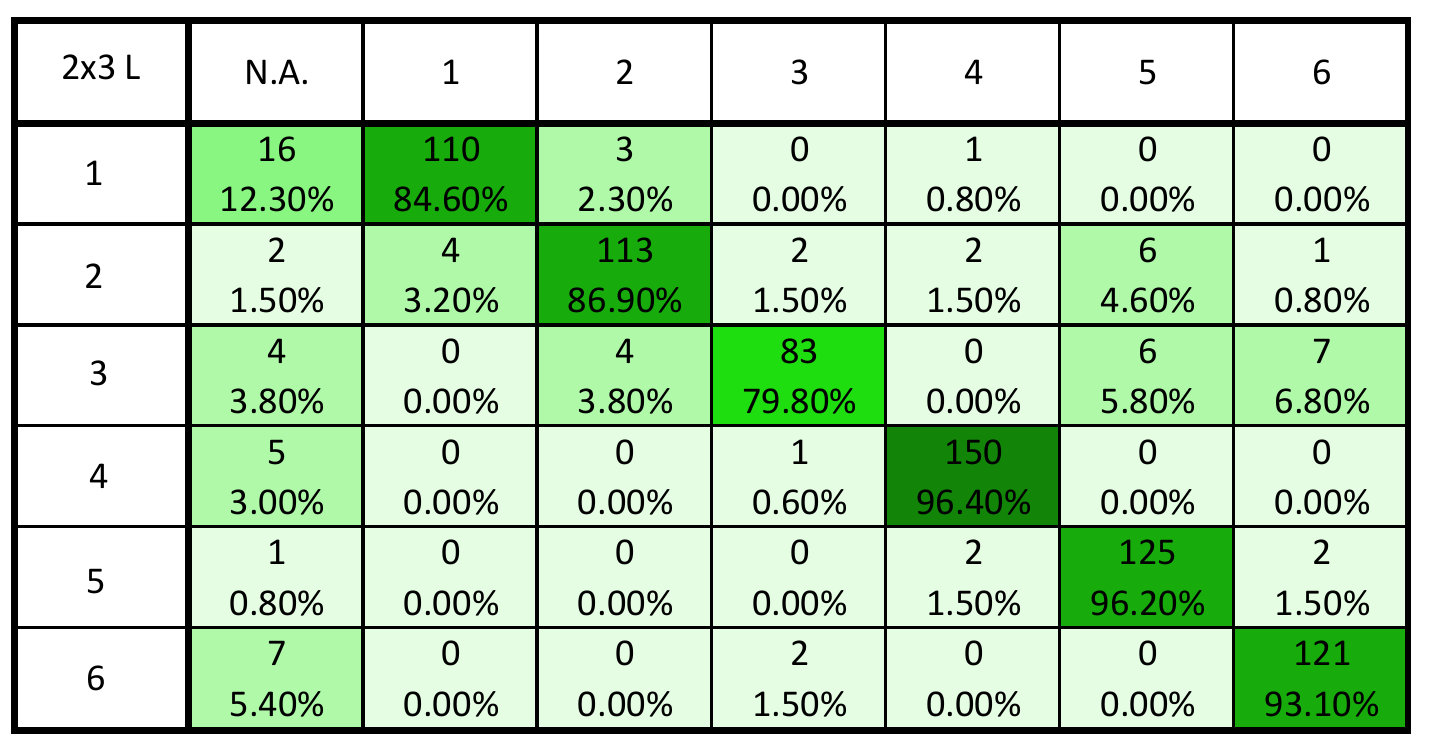}&
      \includegraphics[width=5.5cm,height=4.0cm]{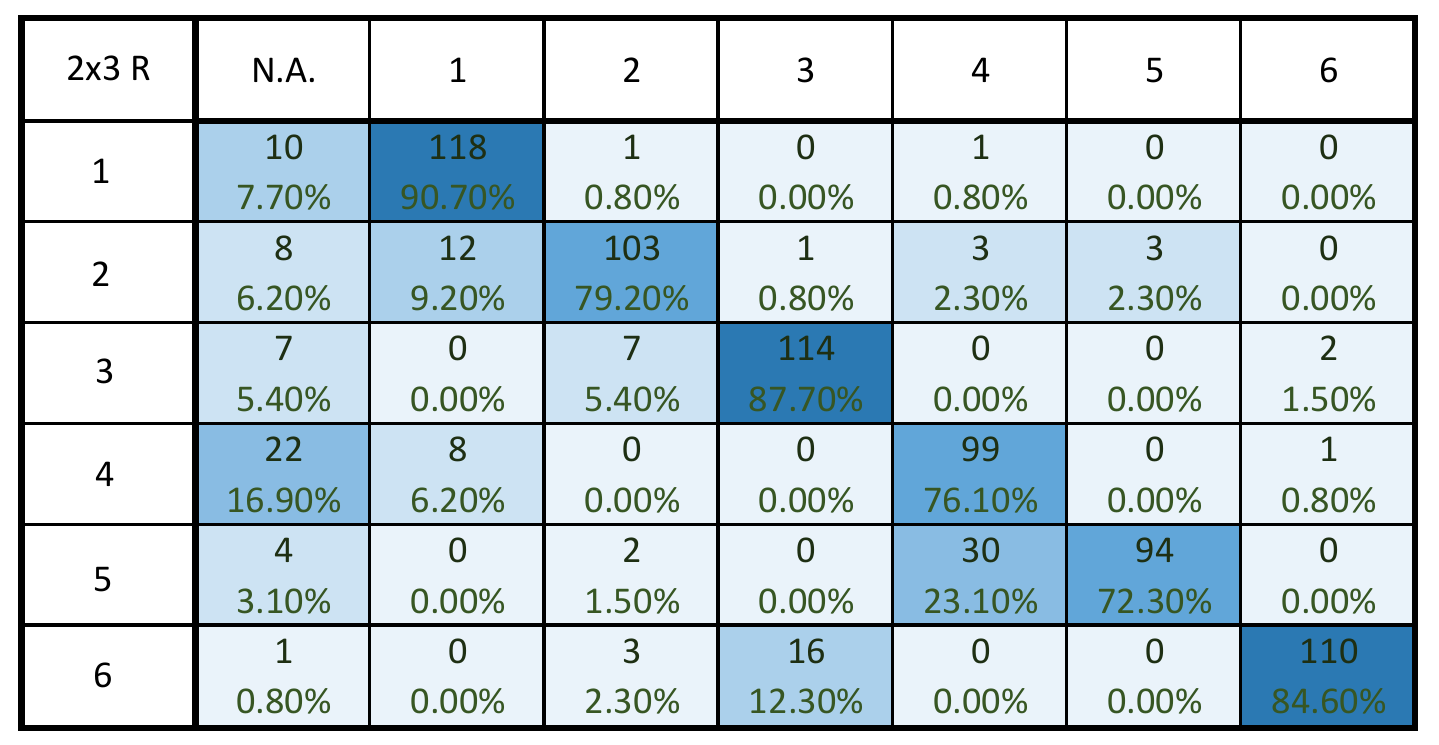}
      \end{tabular}
  \caption{Pointing accuracy for the left/right arm using a $2\times 3$ grid.}
  \label{fig:left23}
 \end{figure}

\begin{figure*}[!h]
 \begin{center}
     \includegraphics[width=12.5cm,height=5.5cm]{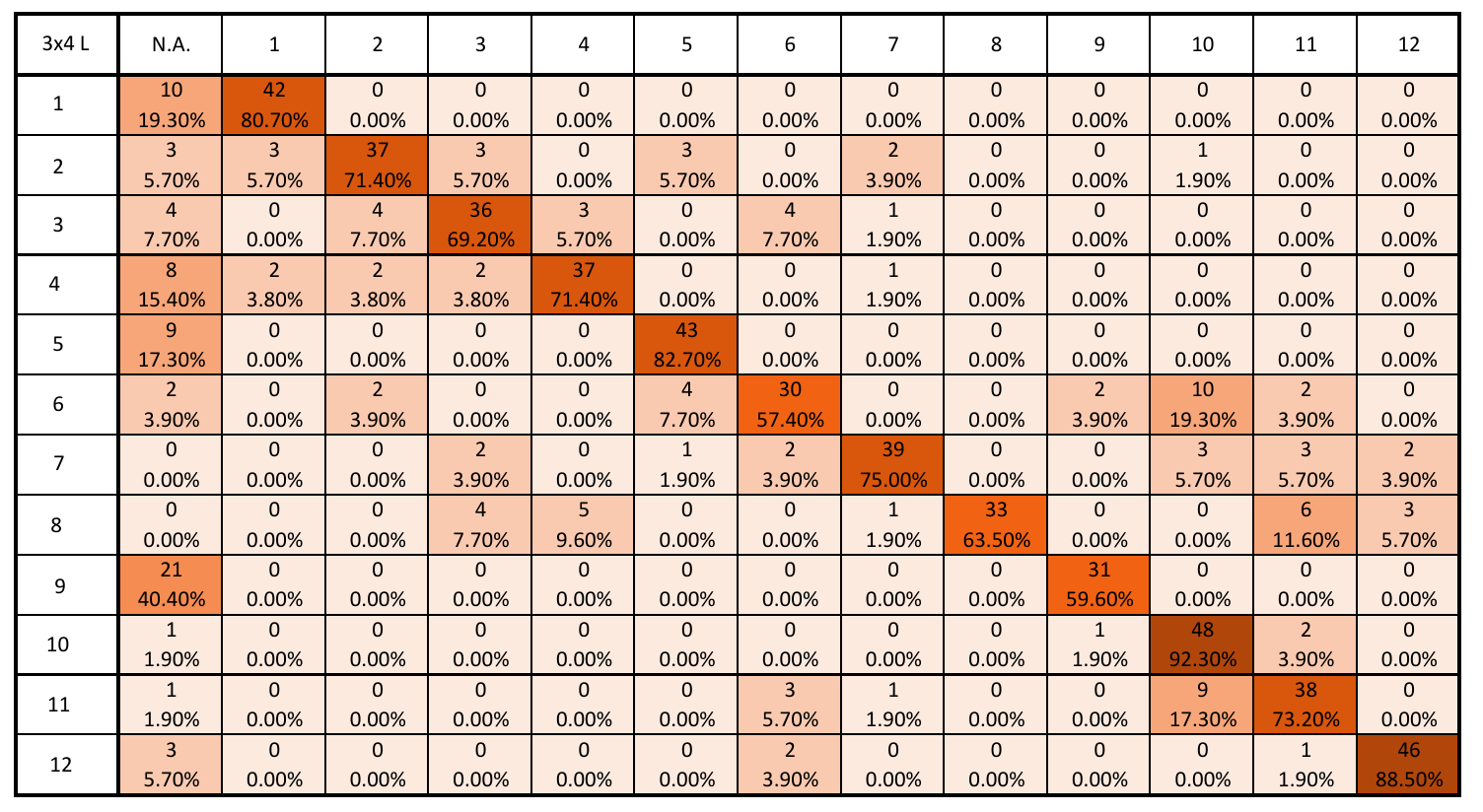}
   \end{center}
  \caption{Pointing accuracy for the left arm using a $3\times 4$ grid.}
  \label{fig:left34}
 \end{figure*}

\begin{figure*}[!h]
   \begin{center}
   \includegraphics[width=12.5cm,height=5.5cm]{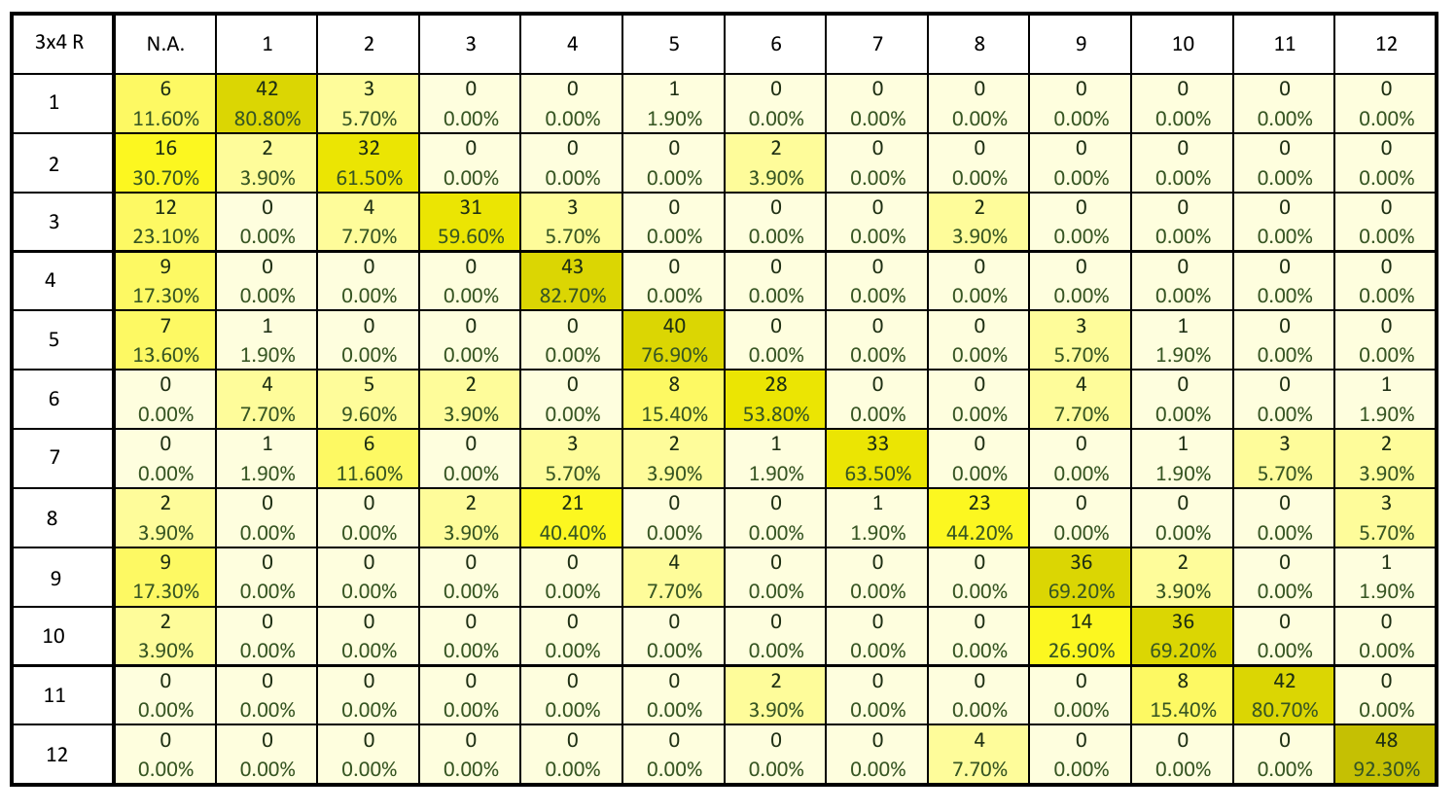}
      \end{center}
  \caption{Pointing accuracy for the right arm using a $3\times 4$ grid.}
    \label{fig:right34}
 \end{figure*}

The system was evaluated in a user study with 26 participants. Different parameters such as handedness, user's height, and arm length were measured. 
Since a user's pointing is significantly influenced by the pointing stability, this also impacts the accuracy of our system, resulting in noticeable differences for the $2 \times 3$ and the $3 \times 4$ grids. The error increases with decreasing box sizes, i.e. it is larger for the $3 \times 4$ grid. The confusion matrix in Figure \ref{fig:left23} left gives an overview on the percentage of the correct pointing at a target number in the $2 \times 3$ matrix using the left arm. Similarly, Figure \ref{fig:left23} (right) describes the quantitative values for right arm corresponding to $2 \times 3$ matrix, Figure \ref{fig:left34} for the left arm pointing at $3 \times 4$ matrix and Figure \ref{fig:right34} for the right arm pointing at $3 \times 4$ matrix.

  Table \ref{tab:accuracy} shows the  accuracy values achieved in the four experiments, i.e., (1) Left arm using $2 \times 3 $ matrix, (2) Right arm using $3 \times 4 $ matrix, (3) Left arm for $3 \times 4 $ matrix and (4) Right arm for $3 \times 4 $ matrix. This accuracy is calculated by converting the output from the system to binary output, i.e., 1, if the output from the system was correct, otherwise it is 0. Then the total number of correct results is divided by the total of trials in the experiment multiplied by 100 to get the percentage of the accuracy.
 
\begin{table}\small
\begin{center}
\begin{tabular}{ccc}
\hline
{} & {Left Arm} & {Right Arm}  \\
\hline
$2 \times 3 $ &  89.59 \% & 79.92 \% \\
$3 \times 4 $  &   73.57 \%  & 68.99 \% \\
\hline

\end{tabular}
\end{center}
\setlength{\belowcaptionskip}{0pt}

\caption{Accuracy for the experiments performed in the user study by using $2 \times 3 $ and $3 \times 4 $ matrix and by using left and the right arm.}
\label{tab:accuracy}
\end{table}

Each of the four tests resulted in a higher accuracy when using the left arm for pointing. This could be caused by the inherent asymmetry within Kinect v2. The IR emitter is centered in the Kinect box, while the IR receiver is off-centered. This leads to a camera's perspective that sees the left arm slightly better than the right one, i.e. the left arm is measured slightly longer than the right one,

\section{\uppercase{Conclusion}}
\label{sec:conclusion}
We worked on automatic pointing gesture detection and pointing target localization in a meeting environment. A prototype of the automatic system was built and tested by conducting a user study. The output of this system will be converted to suitable modality which will help BVIP to get the extra information. Although for our application it is required to have its good performance for 2 x 3 but it proves to have high precision for small areas for localizer function and performs good for both 2 x 3 and 3 x 4 grids in all the four the experiments. We also found out that the stable time for getting the value of localizer is achieved after around 3 seconds and the hand of the user starts to vibrate after an interval again. Our user study also showed that the height of the user did not effect much on the performance. The arm size which is either very small or very large has a small decrease in the accuracy.

In future, the output of our system will be converted by a suitable haptic interface helping BVIP to access these pointing gestures. Moreover, we will expand our system with multiple cameras, and we will have several users pointing simultaneously. Also, we will improve the system to a have more symmetrical output, i.e., the same performance for pointing using the left and the right arm.

\section*{\uppercase{Acknowledgements}}
This work has been supported by the Swiss National Science Foundation (SNF) 
under the grant no. 200021E 177542 / 1. 
It is part of a joint project between
TU Darmstadt,
ETH Zurich,
and JKU Linz 
with the respective funding organizations DFG (German Research Foundation), SNF (Swiss National Science Foundation) 
and FWF (Austrian Science Fund).

%
%
%
 \bibliographystyle{splncs04}
 \bibliography{example}

\end{document}